\documentclass[a4paper,11pt]{article}
\usepackage[utf8]{inputenc}
\usepackage{jheppub}
\usepackage{mathtools,amsmath,amssymb,amsthm,physics}
\usepackage{slashed}
\usepackage{graphicx}
\usepackage{xspace}
\usepackage{subcaption}
\usepackage[table]{xcolor}
\usepackage{tikz-feynman}
\usepackage{nicematrix,tikz}
\usepackage{verbatim}

\renewcommand{\tabcolsep}{2em}

\allowdisplaybreaks

\newcommand{\gaia}[1]{\textbf{\textcolor{orange}{#1}}}
\newcommand{\kay}[1]{\textbf{\textcolor{teal}{#1}}}

\preprint{{ZU-TH 04/24}}

\title{Analytic auxiliary mass flow to compute 
master integrals in singular kinematics}

\author{Gaia Fontana$^{a}$, Thomas Gehrmann$^{a}$, Kay Sch\"onwald$^{a}$}

\affiliation{
         $^a$Physik-Institut, Universit\"at Z\"urich, Winterthurerstrasse 190, 8057 Z\"urich, Switzerland\\
        }

\emailAdd{gaia.fontana@physik.uzh.ch}
\emailAdd{thomas.gehrmann@uzh.ch}
\emailAdd{kay.schoenwald@physik.uzh.ch}

\abstract{
The computation of master integrals from their differential equations requires boundary values 
to be supplied by an independent method. These boundary values are often desired at singular kinematical points. 
We demonstrate how the auxiliary mass flow 
technique can be extended to compute the expansion coefficients 
of master integrals in a singular limit in an analytical manner, thereby providing these boundary conditions. 
To illustrate the application of the method, 
we re-compute the phase space integrals relevant 
to initial-final antenna functions at NNLO, now 
including higher-order terms in their $\epsilon$-expansion in view of their application in 
third-order QCD corrections. 
}

\keywords{Hadron Colliders, QCD Phenomenology, Jets, NNLO Corrections}

\begin{document}
\maketitle

\section{Introduction}
Feynman integrals fulfil differential equations~\cite{Kotikov:1990kg,Remiddi:1997ny} in internal masses and external kinematical 
invariants. These differential equations can be derived at the basis of the integrand, employing integration-by-parts (IBP) reduction~\cite{Chetyrkin:1981qh,Laporta:2000dsw} to express the derivatives 
of the integral under consideration in terms of 
the integral itself plus related inhomogeneous terms. 

Consequently, Feynman integrals can be computed from their differential equations by determining generic solutions that are subsequently matched to boundary values at some special values of masses and kinematics. The resulting particular solutions then produce the desired expressions for the Feynman integrals. The 
differential equation method~\cite{Gehrmann:1999as,Henn:2013pwa} has been 
largely systematised and is now among the most 
powerful and widely used tools to compute 
Feynman integrals, especially in multi-loop, 
multi-scale settings. Variants of it are applied to 
obtain closed-form analytical expressions (often in terms of generalized polylogarithmic functions~\cite{Vollinga:2004sn}) or numerical representations of the integrals. Fairly often, analytical and numerical aspects are combined in hybrid-type approaches. 

The derivation of the differential equations for Feynman integrals  can be largely automated and is
potentially limited only by the complexity and size 
of the 
system of required IBP equations. Once the differential equations are established, they can at least be solved in a purely numerical manner~\cite{Czakon:2008zk} or through a step-wise series expansion algorithm~\cite{Liu:2017jxz,Hidding:2020ytt,Armadillo:2022ugh}, provided their boundary conditions are known. 

Some of these boundary conditions can usually be inferred from internal consistency conditions on the integrals, such as their analyticity behaviour. In most calculations of Feynman integrals with the differential equation method, these considerations are not sufficient to fix all required boundary conditions, thereby necessitating the evaluation of at least some integrals 
for specific kinematical limits by other methods. 

Especially when fully analytical results are desired (as for example to study the transcendental structure of Feynman integrals and Feynman amplitudes), the 
analytical computation of the boundary conditions may become a limiting factor. 

A very promising approach to compute Feynman integrals at fixed kinematical points is the Auxiliary Mass Flow Method (AMFlow,~\cite{Liu:2017jxz}), which employs 
differential equations in an auxiliary mass parameter to determine numerical results for Feynman integrals 
from high-order series expansions to any desired level of numerical precision. 

In this paper, we propose a novel approach to analytically compute Feynman integrals at fixed kinematical points (both regular and exceptional), based on the idea of the AMFlow method of introducing an auxiliary mass parameter. 
Section~\ref{sec:intro-amflow} describes our method and highlights its essential computational steps. We illustrate the application of this method 
in Section~\ref{sec:example} on the analytical computation of boundary values of infrared-divergent phase space integrals relevant to deeply inelastic scattering processes at NNLO, which we compute at their (exceptional) kinematical endpoint. Section~\ref{sec:conc} contains our conclusions and an outlook. 

\section{The Analytic Auxiliary Mass Flow Method}
\label{sec:intro-amflow}
The analytic calculation of loop or phase space integrals in dimensional regularisation is often based on the differential equation method~\cite{Kotikov:1990kg,Remiddi:1997ny,Gehrmann:1999as,Henn:2013pwa}. In this method, differential equations in particle masses or kinematical invariants are derived at the integrand level. By matching 
their generic solutions to  specific boundary conditions, results for the desired integrals are obtained. These boundary conditions are given typically by stating the integrals at specific values of the masses and invariants, obtained from consistency 
conditions or by direct evaluation.

Often there exist special limits in kinematic variables or masses, where the problem simplifies
sufficiently in order to calculate the boundary conditions directly.
For example, in the case of multi-loop 
forward scattering 
amplitudes relevant to 
deep-inelastic scattering the (kinematically unphysical) 
limit of vanishing parton momentum in Euclidian kinematics ($x\to \infty$) can be 
implemented via a naive Taylor expansion in the momentum of the parton and leads 
to simple massless propagator integrals which are known up to four-loop order. This boundary condition is then used in the differential equations in $x$ to obtain results in the physical region $0<x\leq 1$ by analytic 
continuation.
 
For phase space integrals, the dependence on 
external parameters is realized through particle masses 
or by projecting onto a specific final state property such as the momentum fraction or rapidity of an identified final-state particle. 
Especially for phase space integrals, the analytic 
computation of a boundary condition in a 
specific kinematical point is often of comparable 
complexity to the computation of the full integral. 
In particular, those kinematical points that are 
associated with particularly simple expressions usually 
correspond to singular points of the phase space integral, since the external parameters often regulate the infrared behaviour of the integrals under consideration. To circumvent the 
need for a direct computation of boundary conditions 
for phase space integrals, they can  be
obtained in some cases 
indirectly by inclusive integration of the 
generic solution of the differential equations 
over 
a selected external parameter~\cite{Gehrmann:2022cih,Gehrmann:2022pzd}. This approach is however not broadly applicable, since the corresponding inclusive phase space 
integrals may again be of comparable complexity to 
the integrals under consideration. 

The direct evaluation of boundary conditions for phase space integrals usually has to proceed on a process-by-process or even integral-by-integral basis, often requiring dedicated phase-space parametrizations~\cite{Byckling:1971vca} for each integral family. While this can be done for phase-space integrals with rather low multiplicity and 
to low orders in the dimensional regulator $\epsilon$ this procedure becomes 
more and more involved for higher multiplicities and expansion orders.

In the following, we describe an alternative approach to obtain analytic expressions for loop or phase space integrals in specific kinematical points, in particular in singular points. The approach is based on the idea of introducing an auxiliary scale $\eta$ into the integrals under consideration, in such a way that the behaviour in a special limit, usually $\eta \to \infty$, simplifies drastically. In this way one can solve the differential equations with respect to $\eta$, fix the constants of integration, and obtain the original integrals in the limit $\eta \to 0$. This idea has recently been used broadly in the Auxiliary Mass Flow (AMFlow) method~\cite{Liu:2017jxz} for the numerical calculation of loop and phase space integrals. The approach of introducing an auxiliary mass was originally formulated in Ref.~\cite{Schutzmeier:2008sm,Schutzmeier:2009zz} and has been used subsequently in the literature to calculate in particular single-scale integrals which satisfy relatively straightforward differential equations, see e.g.\ Refs.~\cite{Lee:2016lvq,Fael:2020iea,Lee:2023dtc}. Its implementation in the form of the AMFlow method allows to address a much larger range of multi-scale problems in a largely automated manner, yielding results in high-precision numerical form.

The original formulation~\cite{Schutzmeier:2009zz,Liu:2017jxz} of the method can be summarized as follows.
One considers an integral family characterized by the set of denominators $\{ D_i \}$
and the corresponding master integrals $\vec{M}(\epsilon,s_{ij})$ which are functions of the space-time dimension 
$d=4-2\epsilon$ and kinematic invariants $s_{ij}$.
This  original integral family is 
modified by adding an auxiliary mass $\eta$ to each propagator 
of the integral family 
\begin{align} 
    D_i \to D_i - \eta^2 
    ~.
\end{align} 
By adding the auxiliary mass $\eta$ one obtains a new set of master integrals $\vec{M}_\eta(\epsilon,s_{ij},\eta)$. 
Due to the increased complexity induced by the massive propagators the number of master integrals is usually larger than before. 
The new set of master integrals fulfils a differential equation in the auxiliary mass 
\begin{align} 
    \frac{d}{d \eta} \vec{M}_\eta(\epsilon,s_{ij},\eta) &= A(\epsilon,s_{ij},\eta) \cdot \vec{M}_\eta(\epsilon,s_{ij},\eta)
    ~.
\end{align}
For these integrals, one  can calculate the limit $\eta \gg s_{ij}$ by using the large mass expansion~\cite{Smirnov:1990rz}
which is a special case of the expansion-by-regions method~\cite{Beneke:1997zp}.
Since all propagators are massive, only the region where all loop momenta $l_i$ scale proportional to the 
heavy mass ($l_i \to \eta l_i$) contributes to the expansion.
Therefore the integrals for $\eta\to \infty$ reduce to fully massive tadpole integrals which are known analytically 
up to three-loop order \cite{Davydychev:1992mt,Broadhurst:1998rz,Kniehl:2017ikj} and to very high precision numerically up to five-loop order
\cite{Schroder:2005va,Luthe:2015ngq,Luthe:2017ttc}.
One can then use the differential equation to compute a deep series solution around $\eta \to \infty$,
employing the $\eta\to \infty$ behaviour to determine all constants of integration.
Finally, one can transport the initial value from large values of $\eta$ to the physical values of $\eta \to 0$ by 
constructing consecutive overlapping series expansions and matching them at points where two neighbouring expansions converge.
Typically, the limit $\eta \to 0$ is not analytic, but also described by an asymptotic expansion, so 
one has to be careful when taking the limit $\eta \to 0$.

The drawback of this approach is that adding masses to all propagators increases 
the number of master integrals tremendously, which can render the reduction problem 
or the numerical solution of the master integrals unfeasible.
In order to remedy this problem the method has been refined in Ref.~\cite{Liu:2021wks,Liu:2022mfb}.
In the refined procedure, the auxiliary mass is not necessarily  added to all propagators, but one 
chooses a smaller set of propagators to add the mass to. 
This allows to control the number of master integrals.
The refinement comes with its own drawbacks, since 
 the asymptotic expansion for $\eta \to \infty$ becomes more involved with more than one region 
 potentially contributing in this limit. Moreover, 
 the limiting values at $\eta \to \infty$  are not fully massive tadpoles anymore. 
The first point can still be handled algorithmically by the procedure of the large mass expansion, 
however the complexity increases because of the larger number of regions which have to be considered.
The second point can be handled by applying the AMFlow method again on the $\eta\to \infty$ integrals, therefore reducing the problem to fully massive tadpoles recursively~\cite{Liu:2021wks,Liu:2022mfb}.

The method has also been extended to handle phase-space integrals~\cite{Liu:2020kpc}.
Here the basic idea is to use reverse unitarity \cite{Anastasiou:2002yz,Anastasiou:2003gr} to treat the phase-space integrals 
on the same footing as loop integrals by replacing phase-space constraints with cut propagators. 
When applying the AMFlow method, the auxiliary mass is not added to the cut propagators.
In this way one can use the AMFlow method to recursively reduce the boundary integrals to 
precomputed and simple phase-space integrals. 

The methods described above have been implemented in a public package \texttt{AMFlow}~\cite{Liu:2022chg}.
Since the method increases the complexity of the integrals and therefore also 
of the reduction problem and the differential equations, an essential point of the 
implementation is that all kinematic variables $s_{ij}$  and the dimensional regulator $\epsilon$  are 
set to rational values right from the beginning. The $\epsilon$-dependence of the result is subsequently 
reconstructed by fitting to an appropriate ansatz. 
While this is useful to speed up numerical calculations it has the drawback that the 
resulting integrals 
cannot be obtained analytically. It also 
restricts the application of the method to 
non-exceptional kinematical points, since 
 possible singular regions of the asymptotic expansion  implicitly vanish 
in dimensional regularization once the variable is set to the singular value and one can only recover 
the so-called hard region of the asymptotic expansion.
However, Feynman integrals usually simplify considerably in singular limits, making them 
ideal candidates for the calculation of boundary conditions to the differential equation method. 

To apply the auxiliary mass flow method to 
the computation of integrals in 
singular points requires to retain the analytic 
dependence on the parameter that characterizes the 
singular limit. The resulting Analytic Auxiliary Mass 
Flow method (AAMFlow) is described in the following. 

To make the discussion easier we consider an integral 
depending on one parameter $y$. The generic solution of the differential equation in $y$ is known. To obtain the particular
solution for the integral, we thus 
require the boundary condition, which we 
aim to calculate in the  singular limit $y \to 0$. 
After adding the auxiliary mass we have the differential equation for the master integrals 
\begin{align}
    \frac{d}{d \eta} \vec{M}_\eta(\epsilon,y,\eta) &= A(d,y,\eta) \cdot \vec{M}_\eta(\epsilon,y,\eta)
    ~.
    \label{eq::deq}
\end{align}
In the limit $y \to 0$ we can use the reasoning of expansion-by-regions to find an expansion of the form 
\begin{align} 
    M_{\eta}^{(l)} &= \sum_{i} y^{n_i^{(l)} + d_i \epsilon} \sum\limits_{j=0}^{\infty} c_{ij}^{(l)}(\eta,\epsilon) y^j
    \,,
    \label{eq::ansatz}
\end{align}
where the superscript $(l)$ enumerates the individual 
components of the vector of master integrals.
 There is only a finite number of $d_i$ to consider and $d_i=0$ corresponds to the so-called hard region.

The starting powers $n_i^{(l)}$ are specific for each region and master integral and can usually be determined 
beforehand or at least a lower bound can be extracted from physical arguments.
We can now insert the ansatz~\eqref{eq::ansatz} into the differential equation~\eqref{eq::deq}
and obtain differential equations for the expansion coefficients $c_{ij}^{(l)}(\eta,\epsilon)$.

Since we are only interested in the leading terms of the asymptotic expansion (i.e. $j=0$)
to find boundary conditions for the differential equation for general $y$, we only have to consider 
a rather small set of expansion coefficients which we denote with $\vec{c}(\eta,\epsilon)$.  
It is worth mentioning that the different branches of the asymptotic expansion, which are characterized by 
different $d_i$, do not mix in the differential equation with respect to $\eta$.
We therefore find simplified differential equations 
\begin{align} 
    \frac{d}{d \eta} \vec{c}_i &= \hat{A}_i(\eta,\epsilon) \cdot \vec{c}_i 
    ~, 
    \label{eq::deqC}
\end{align}
where the subscript $i$ labels the factorizing branches of the differential equation.
We observe that the differential equations for the coefficients $\vec{c}_i(\eta,\epsilon)$ simplify 
since they do not depend on $y$ anymore. 

However, some care is required in selecting  those propagators that the auxiliary mass $\eta$ is added 
to. The addition of $\eta$ should not modify the singular behaviour of the integral in $y \to 0$, 
since the expansions in $\eta \to 0$ and $y \to 0$ are used interchangeably and therefore have to commute. 
This commutativity can be validated at the level of 
the differential equations, which must not contain 
denominator factors
that are finite for  $\eta \to 0$ or $y \to 0$ 
but vanish for the simultaneous limit 
($\eta \to 0$ and $y \to 0$). 
A targeted choice of propagators with auxiliary mass 
allows to avoid these pathological situations. 

At this point,  the expansion coefficients $\vec{c}_i(\eta,\epsilon)$ fulfil a differential equation with respect to $\eta$ 
and the constants of integration can be fixed in the limit $\eta \to \infty$. Consequently,  
the $y\to 0$ limit of the original master integrals
can now be calculated numerically to high precision
using the original AMFlow method. 

However, the reduced complexity of the differential equation in $\eta$ might allow an analytic treatment
with the established techniques of multi-loop calculations.
For example one can try to find a transformation to a canonical differential equation \cite{Henn:2013pwa},
i.e.\ a new basis $\vec{\tilde{c}}_i$ in which the differential equation \eqref{eq::deqC} takes the form 
\begin{align} 
    \frac{d}{d \eta} \vec{\tilde{c}}_i(\eta,\epsilon) &= \epsilon \tilde{A}_i(\eta) \cdot \vec{\tilde{c}}_i(\eta,\epsilon)
    ~,
\end{align} 
or to find solutions in terms of iterated integrals following the algorithms outlined in~\cite{Ablinger:2018zwz}.
Once closed-form solutions for the $\vec{\tilde{c}}_i(\eta,\epsilon)$ are found, the limit $\eta \to 0$ has to be taken.
As has been mentioned before, this limit is also described by an asymptotic expansion. Taking 
into account that each loop momentum can scale 
in a soft or a hard manner relative to $\eta$, we 
 find the following ansatz for the limit $\eta \to 0$
\begin{align} 
    c_{ij}^{(l)}(\eta,\epsilon) = \sum_{\ell=0}^{L} \eta^{-\ell\epsilon} d_{ij,\ell}^{(l)}(\epsilon) + \mathcal{O}(\eta) 
    \label{eq:flow0}
    ~,
\end{align}
where $L$ corresponds to the number of (un-cut or cut) loops.
The physical boundary conditions that we aim for  
are given by the hard region ($\ell=0$).
One way to obtain these physical results is therefore to expand the solution to sufficiently high orders 
in the dimensional regulator and use the knowledge of terms proportional to powers of $\log(\eta)$ to 
disentangle the regions.

For the two loop integrals considered in this paper, the ansatz for the $\eta \rightarrow 0$ limit is explicitly given by
\begin{align}
      c_{ij}^{(l)}(\eta,\epsilon) = d_{ij,0}^{(l)}(\epsilon) + \eta^{-\epsilon} d_{ij,1}^{(l)}(\epsilon) + \eta^{-2\epsilon} d_{ij,2}^{(l)}(\epsilon) + \mathcal{O}(\eta) 
\end{align}
where the expansion coefficients $d_{ij,\ell}^{(l)}(\epsilon)$ still have a Laurent expansion in the regulator $\epsilon$. 
We indicate their $\epsilon-$expansion as:
\begin{align} 
d_{ij,R}^{(l)} = \sum_{k=\text{min}}^{\infty} \epsilon^k d_{ij,R}^{(l),k}, \quad R=0,1,2.
\end{align}
At each order in $\epsilon$ the coefficients $c_{ij}^{(l)}(\eta,\epsilon)$ can be written as
\begin{align}
    c_{ij}^{(l)}(\eta,\epsilon) = \sum_{k=\text{min}}^{\infty} 
    \epsilon^k \biggl[ r_{k,0} + \sum\limits_{m=1}^k r_{k,m} \log^m(\eta) \biggr]
    ~,
\end{align}
leaving implicit the dependence of $ r_{k,m} \,(m\geq 0)$ on $i,j,\ell$.
We have
\begin{align}
   r_{k,0} =  d_{ij,0}^{(l),k} + d_{ij,1}^{(l),k} + d_{ij,2}^{(l),k} 
\end{align}
and $r_{k,m} \,( m\geq 1)$ representing the coefficients of the $\log(\eta)$ terms in the expansion. It is crucial to notice that once the integral has been solved, all these coefficients are known, and it is trivial to extract each $r_{i,k}$. Our goal is to obtain the expression for the hard region, specifically each $d_{ij,0}^{(l),k}$. 

We demonstrate how this can be done by looking at the first coefficient, $d_{ij,0}^{(l),0}$, assuming without loss of generality that the hard-region expansion starts at order $\mathcal{O}(1)$. We can extract $d_{ij,0}^{(l),0}$ by examining the analytic structure of the $\epsilon$-expansion of the $c_{ij}^{(l)}(\eta,\epsilon)$, recognizing which terms contribute to it.

We expand the ansatz in $\epsilon$ and we obtain up to $\mathcal{O}(\epsilon^3)$
\begin{align}
c_{ij}^{(l)} =
& \,\, d_{ij,0}^{(l),0}+d_{ij,1}^{(l),0}+d_{ij,2}^{(l),0} \nonumber\\
+&\, \epsilon \, \left(d_{ij,0}^{(l),1}+\left(-d_{ij,1}^{(l),0}-2 d_{ij,2}^{(l),0}\right) \log (\eta)+d_{ij,1}^{(l),1}+d_{ij,2}^{(l),1}\right)\nonumber\\
+&\, \epsilon^2 \, \left(d_{ij,0}^{(l),2}+d_{ij,1}^{(l),2}+d_{ij,2}^{(l),2}+\frac{1}{2} \left(d_{ij,1}^{(l),0}+4 d_{ij,2}^{(l),0}\right) \log ^2(\eta)+\left(-d_{ij,1}^{(l),1}-2 d_{ij,2}^{(l),1}\right) \log (\eta)\right)\nonumber\\
+\, & \mathcal{O}(\epsilon^3).
\end{align}
The $\mathcal{O}(1)$ term is given by the sum of the leading coefficients of the three regions. To extract the hard-region, our strategy will consist in subtracting from the $\mathcal{O}(1)$ term the sum of the leading term of the $\eta^{-\epsilon}$ and $\eta^{-2\epsilon}$ regions, respectively $d_{ij,1}^{(l),0}$ and $d_{ij,2}^{(l),0}$. These two terms can in turn be determined by the coefficients of the $\log(\eta)$ terms in the higher order terms of the expansion. We can see that at $\mathcal{O}(\epsilon)$ and $\mathcal{O}(\epsilon^2)$ we have that the coefficient of $\log(\eta)$ and of $\log^2(\eta)$ are given by
\begin{align}
\begin{cases}
    -d_{ij,1}^{(l),0}-2 d_{ij,2}^{(l),0} = r_{1,1} ~,\\
    d_{ij,1}^{(l),0} / 2+2 d_{ij,2}^{(l),0} = r_{2,2} ~, \\
     d_{ij,0}^{(l),0} + d_{ij,1}^{(l),0} + d_{ij,2}^{(l),0} = r_{0,0} ~.
\end{cases}
\end{align}
This linear system of equations can be solved in terms of $d_{ij,0}^{(l),0}$ easily.

The outlined procedure can be easily extended to fix any coefficient $d_{ij,0}^{(l),k}$ by first fixing the $d_{ij,1}^{(l),k}$ and $d_{ij,2}^{(l),k}$ by solving the system 
\begin{align}
\begin{cases}
    -d_{ij,1}^{(l),k}-2 d_{ij,2}^{(l),k} = r_{k+1,1} ~,\\
    d_{ij,1}^{(l),k} / 2+2 d_{ij,2}^{(l),k} = r_{k+2,2} ~,\\
    d_{ij,0}^{(l),k} + d_{ij,1}^{(l),k} + d_{ij,2}^{(l),k} = r_{k,0} ~,
\end{cases}
\label{eq:fix}
\end{align}
and then subtracting their sum from the coefficient of order $\epsilon^k \log(\eta)^0$.

In the following sections,  we will illustrate the 
application of this method on the analytical 
computation of phase space integrals relevant to 
double real radiation and to single real radiation 
at one loop, both in the kinematics of deeply 
inelastic scattering (DIS). 
Integrals of this type were first 
considered in the context of the NNLO QCD corrections 
to DIS structure functions~\cite{Zijlstra:1992qd}. 
In the context of the development of the antenna subtraction 
method~\cite{Gehrmann-DeRidder:2005btv,Daleo:2006xa,Currie:2013vh} at NNLO, 
the full set of phase space master integrals was computed~\cite{Daleo:2009yj} through to finite terms in $\epsilon$, yielding the integrated NNLO antenna functions in initial-final kinematics. In view of extending the antenna subtraction method to 
N3LO~\cite{Jakubcik:2022zdi,Chen:2023fba,Chen:2023egx}, 
we re-compute these integrals to two more orders in 
their expansion in $\epsilon$, thus providing the 
integrated NNLO antenna functions 
to order $\epsilon^2$.

The phase space master integrals depend on the kinematical scaling parameter $z$, and we compute them from their differential equations. 
The method outlined above is applied to find the boundary values in the singular soft limit $z\to 1$ systematically up to transcendental weight 6, as required for the subleading terms in the $\epsilon$-expansion of the integrated antenna functions. 

\section{Example application: phase space master integrals}
\label{sec:example}
In the following, we will describe in detail how the 
analytic auxiliary mass flow method (AAMFlow) 
can be applied to determine the analytic values 
of NNLO phase space master integrals relevant
to DIS in their singular kinematical endpoint $z\to 1$. 
These values are then used as boundary conditions 
to differential equations for these master integrals, thereby enabling their analytic computation. 

\subsection{Setup}

\subsubsection{Kinematics and notation}

The NNLO real radiation corrections to inclusive 
DIS contain contributions from double real radiation (RR) processes and from the one loop virtual corrections to 
single real radiation (real-virtual, RV) processes. 
Their kinematics is described by 
\begin{equation}
    q_1+ q_2 \rightarrow p_1 + p_2 (+ p_3),
    \label{eq:kinem}
\end{equation}
where
\begin{align}
    q_1^2=0, \quad q_2^2=-Q^2<0, \quad p_i^2=0 \quad (i=1,2,3),
\end{align}
and $p_3$ is only present for the RR processes. 

The initial state kinematics is fully described 
by the dimensionful $Q^2$ and a dimensionless scaling variable $z$:
\begin{equation}
    z \equiv \frac{Q^2}{2 q_1\cdot q_2},
\end{equation}
related to the Mandelstam invariant $s=(q_1+q_2)^2$ as 
\begin{equation}
    s = Q^2 \Big( \frac{1-z}{z} \Big).
    \label{eq:sandzrel}
\end{equation}
In the following we fix $Q^2=1$, since dimensional analysis allows us to reconstruct the dependence on $Q^2$ at a later stage.

The integrated initial-final NNLO antenna functions (and likewise, the DIS coefficient functions) 
are 
obtained by integrating fully inclusively over 
the final state phase space. The RR and RV integrands 
can be expressed in terms of a properly chosen set 
of linearly independent propagator factors $D_j$ as:
\begin{equation}
    I_{RR} = \int \, \dd \Pi_3 \,(2\pi)^d\, \delta^{(d)}\left(q_1 + q_2 - \sum_{i=1}^3 p_i\right) \prod_{j}\frac{1}{D_j^{\alpha_j}}, \quad \alpha_j \in \mathbb{Z}
\end{equation}
and 
\begin{equation}
    I_{RV} = \int \, \dd \Pi_2 \, (2\pi)^d \, \delta^{(d)}\left(q_1 + q_2 - \sum_{i=1}^2 p_i\right) \int \frac{\dd^d k}{(2\pi)^d} \prod_{j}\frac{1}{D_j^{\alpha_j}},\quad \alpha_j \in \mathbb{Z}. 
\end{equation}
The vectors $v_j$ that constitute the propagator factors $D_j=v_j^2$ are constructed from linear combinations of the $q_i,p_i$ and, in the RV case, of the loop momentum $k$. They form a complete linearly independent basis, allowing to express any scalar product that appears in the numerator of the integrand.  

The measure $\dd \Pi_n$ for the $n$-particle phase-space reads
\begin{equation}
    \dd \Pi_n =  \prod_{i=1}^n \frac{\dd^d p_i}{(2\pi)^{d-1}} \delta^+(p_i^2), \quad \delta^+(p_i) = \delta(p_i^2) \theta(p_i^0). 
\end{equation}
To calculate phase-space integrals we employ the \emph{reverse unitarity} method~\cite{Anastasiou:2003gr,Anastasiou:2002yz}, which allows us to map phase-space integrals to loop integrals with forward kinematics and cut propagators, 
following from the Cutkosky rules \cite{Cutkosky:1960sp}:
\begin{equation}
    \frac{1}{(\slashed{p_i}^2)} = - 2 \pi i \delta^+(p_i^2) = \frac{1}{p_i^2+ i0} - \frac{1}{p_i^2- i0},
\end{equation} 
where we indicate the cut of a generic propagator $D_i$ as
\begin{equation}
   \slashed{D}_i \equiv (D_i)_{\text{cut}}. 
\end{equation}

Using reverse unitarity, the RR-integrals therefore
have the following structure 
\begin{equation}
    I_{RR} = (-i)^3 \,\int \prod_{i=1}^3 \left(\frac{\dd^d p_i}{(2 \pi)^d}\frac{1}{\slashed{p_i}^2}\right)\,(2 \pi)^d\, \delta\left(q_1 + q_2 - \sum_{i=1}^3 p_i\right) \prod_{j \in \text{uncut}}\frac{1}{D_j^{\alpha_j}}, \quad \alpha_j \in \mathbb{Z}.
\end{equation}
 On the other hand, the RV-integrals have the form
\begin{equation}
    I_{RV} = (-i)^2\,\int \prod_{i=1}^2 \left( \frac{\dd^d p_i}{(2 \pi)^d}\frac{1}{\slashed{p_i}^2} \right)\,(2\pi)^d\, \delta\left(q_1 + q_2 - \sum_{i=1}^3 p_i\right) \int \frac{\dd^d k}{(2\pi)^d} \prod_{j \in \text{uncut}}\frac{1}{D_j^{\alpha_j}}, \, \alpha_j \in \mathbb{Z}.
\end{equation}
With this setup, both RR and RV phase space integrals are mapped to cuts of two-loop four-point integrals in forward kinematics. 

The map between phase-space integrals and cut-loop integrals allows to use well-established techniques for the evaluation of multi-loop integrals. In particular, we reduce any RR or RV integral to a linear combination of master integrals~\cite{Daleo:2009yj} by using integration-by-parts (IBP) identities~\cite{Tkachov:1981wb,Chetyrkin:1981qh,Laporta:2000dsw}. The master integrals are subsequently computed from their differential equations~\cite{Kotikov:1990kg,Remiddi:1997ny,Gehrmann:1999as}.

\subsubsection{Integral families}

The RR and RV phase-space integrals can be obtained as the physical three-particle and two-particle  cuts of the two-loop forward scattering amplitude. The loop integrals for the latter were classified in~\cite{Blumlein:2022gpp}, and we derive the RR and RV integral families by identifying all their topologically distinct three-particle and two-particle cuts. 

All integrals can be expressed in terms of linearly independent subsets (integral families), each containing 7 elements, of the following 14 propagators:
\begin{align}
    D_1 &= k_1^2 \nonumber \\
    D_2 &= k_2^2 \nonumber \\
    D_3 &= (k_1-q_1)^2 \nonumber \\
    D_4 &= (k_2+q_1)^2 \nonumber \\
    D_5 &= (k_1-k_2-q_1)^2 \nonumber \\
    D_6 &= (k_1+q_2)^2 \nonumber \\
    D_7 &= (k_1-k_2-q_1+q_2)^2 \nonumber \\
    D_8 &= (k_2-q_1)^2 \nonumber \\
    D_9 &= (k_2+q_2)^2 \nonumber \\
    D_{10} &= (k_1+k_2-q_1+q_2)^2 \nonumber \\
    D_{11} &= (k_1-q_2)^2 \nonumber \\
    D_{12} &= (k_2-q_2)^2 \nonumber \\
    D_{13} &= (k_1+k_2-q_1-q_2)^2 \nonumber \\
    D_{14} &= (k_1+k_2)^2.
    \label{eq:props}
\end{align}
All RR integrals must contain exactly 3 cut propagators. They can be mapped to the following four integral families:
\begin{align}
    \text{A}&= \left\{ D_1, \slashed{D}_2, D_3, D_4, \slashed{D}_5, \slashed{D}_6, D_7\right\} ,\nonumber\\
    \text{B}&=\left\{ D_1, D_2, D_3, \slashed{D}_6, D_8, \slashed{D}_9, \slashed{D}_{10} \right\}, \nonumber\\
    \text{C}&=\left\{ D_1, D_2, \slashed{D}_3, D_6, \slashed{D}_8, D_9, \slashed{D}_{10}\right\} ,\nonumber\\
    \text{D}&=\left\{\slashed{D}_1, \slashed{D}_2, D_3, D_8, D_{11}, D_{12}, \slashed{D}_{13}\right\}.
    \label{eq:famRR}
\end{align}

\begin{figure}[t]
     \centering
     \begin{subfigure}[b]{0.3\textwidth}
         \centering
         \includegraphics[width=\textwidth]{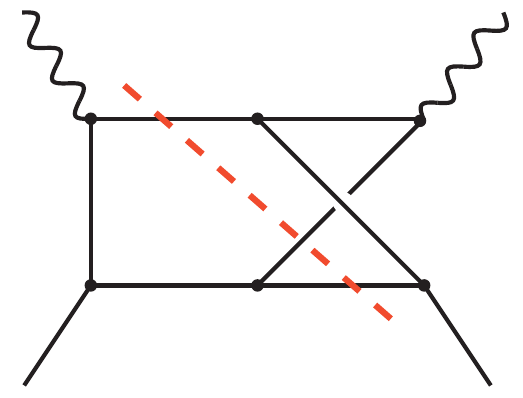}
         \caption{Integral family A.}
         \label{fig:famA}
     \end{subfigure}
     \hspace{0.1\textwidth}
     \begin{subfigure}[b]{0.3\textwidth}
         \centering
         \includegraphics[width=\textwidth]{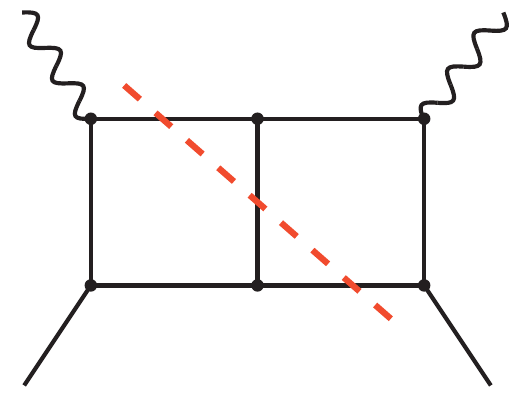}
         \caption{Integral family B.}
         \label{fig:famB}
     \end{subfigure}
     \bigskip
     \begin{subfigure}[b]{0.3\textwidth}
         \centering
         \includegraphics[width=\textwidth]{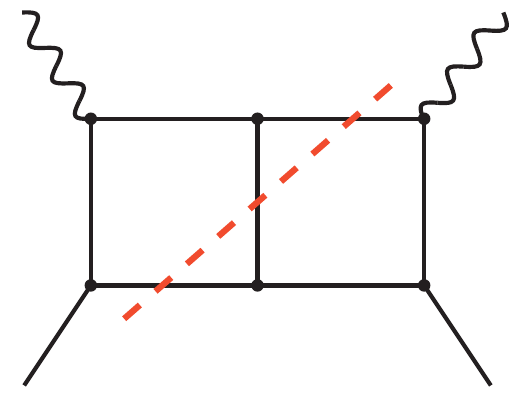}
         \caption{Integral family C.}
         \label{fig:famC}
     \end{subfigure}
     \hspace{0.1\textwidth}
      \begin{subfigure}[b]{0.35\textwidth}
         \centering
         \includegraphics[width=\textwidth]{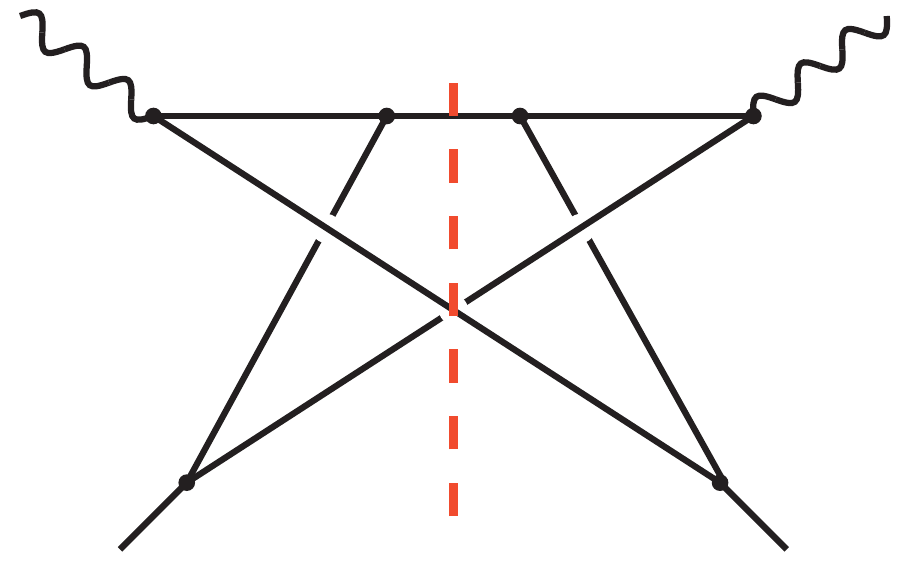}
         \caption{Integral family D.}
         \label{fig:famD}
     \end{subfigure}
        \caption{Top sectors of integral families A,B,C,D.}
        \label{fig:fams}
\end{figure}

Figure~\ref{fig:fams} depicts the top sector integral (integral with three cut and four uncut propagators) of each sector, which can not have any irreducible scalar products in the numerator due to combinatorics. 

The reduction of all integrals in a given family is performed using \texttt{Kira}~\cite{Klappert:2020nbg}. 
The list of RR master integrals, taking into account the symmetry relations between the families is
\begin{align}
    &I_0,\nonumber\\
    &I^A_{1},  I^A_{1,7},I^A_{1,7^2}, I^A_{3,7}, I^A_{3,4,7},\nonumber \\ &I^B_{1,2,3,8},\nonumber\\
    &I^C_{1,2,6,9},\nonumber\\
    &I^D_{3,8,11,12},
\end{align}
where we used the notation $I^{\text{family}}_{\text{list of uncut propagators}}$, with superscripts in the list indicating propagators raised to higher powers, and $I_0$ corresponds to the integration of the three-particle phase-space measure without any propagators~\cite{Gehrmann-DeRidder:2003pne}:
\begin{equation}
    I_0 = S_\Gamma e^{2\gamma_E\epsilon} \left(\frac{1-z}{z}\right)^{1-2\epsilon} \frac{2 \pi \Gamma(1-\epsilon)^3}{\Gamma(3-3\epsilon)\Gamma(2-2\epsilon)}~,
\end{equation}
where we choose the normalization 
\begin{equation}
    S_\Gamma = \left( \frac{(4\pi)^\epsilon e^{-\gamma_E\epsilon}}{16\pi^2} \right)^2 ~.
\end{equation}

\begin{figure}[t]
     \centering
     \begin{subfigure}[b]{0.3\textwidth}
         \centering
         \includegraphics[width=\textwidth]{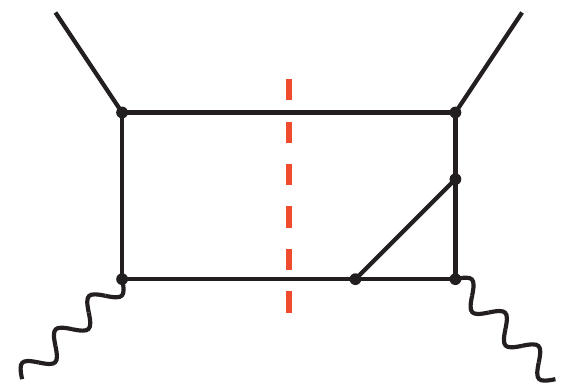}
         \caption{Integral family E.}
         \label{fig:famE}
     \end{subfigure}
     \hspace{0.1\textwidth}
     \begin{subfigure}[b]{0.3\textwidth}
         \centering
         \includegraphics[width=\textwidth]{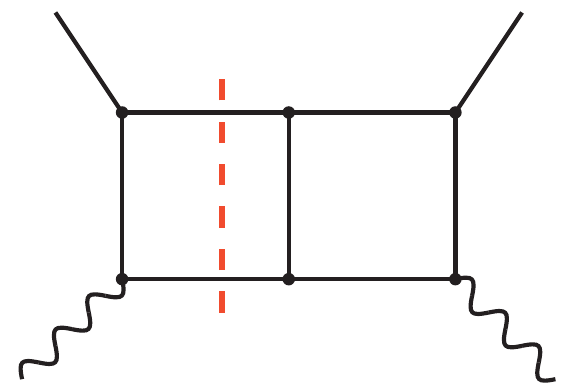}
         \caption{Integral family F.}
         \label{fig:famF}
     \end{subfigure}
     \hspace{0.1\textwidth}
     \begin{subfigure}[b]{0.3\textwidth}
         \centering
         \includegraphics[width=\textwidth]{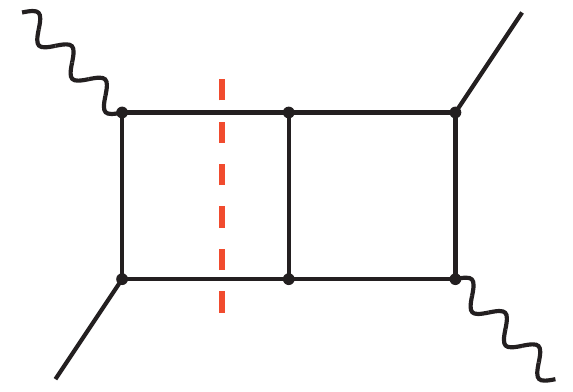}
         \caption{Integral family G.}
         \label{fig:famG}
     \end{subfigure}
        \caption{Top sectors of integral families E,F,G.}
        \label{fig:famsRV}
\end{figure}

All RV integrals must contain exactly 2 cut propagators. They can be mapped to the following three integral families, displayed in figure~\ref{fig:famsRV}:
\begin{align}
    \text{E}&= \left\{ D_1, D_2 , \slashed{D}_3, D_4, D_5 , \slashed{D}_6, D_7\right\} ,\nonumber\\
    \text{F}&=\left\{ D_1, D_2, \slashed{D}_3, D_4, \slashed{D}_6, D_{12}, D_{14}\right\}, \nonumber\\
    \text{G}&=\left\{D_1, D_2, \slashed{D}_3,\slashed{D}_6, D_8, D_9,D_{10}\right\}.
    \label{eq:famRV}
\end{align}
The reduction of all integrals in a given family is again performed using \texttt{Kira}~\cite{Klappert:2020nbg}. The list of RV master integrals, taking into account the symmetry relations between the families is
\begin{align}
    &I^E_{4,5},  I^E_{5,7},I^E_{2,5,7}, I^E_{2,4,5,7},\nonumber \\ &I^F_{4,12},\nonumber\\
    &I^G_{1,2,8,9,10}
    ~.
\end{align}
These bases differ from the one presented in~\cite{Daleo:2009yj}. The explicit transformations are given by
\begin{align}
  I[0] =&I_0, \qquad I[2] = I^A_1, \qquad I[2,6] = I^A_{3,7}, \qquad I[2,4,9] = I^A_{3,4,7}~, \nonumber\\
  I[1,2,5] =& \frac{3 (d-3) (3 d-10) z^2 }{(d-4)^2 (z+1)}\,I^A_1-\frac{2 z }{(d-4) (z+1)}\,I^A_{1,7^2}~,\nonumber\\
  I[2,3,5] =& -\frac{2 (3 d-11) z^2 }{(d-4) (z-1)^2}\,I^A_{1,7}-\frac{2 (d-3) (3 d-10) z^2 }{(d-4)^2 (z-1)^2}\,I^A_1 \nonumber\\
  &+\frac{2 z (z+1) }{(d-4) (z-1)^2}\,I^A_{1,7^2}+\frac{z }{z-1}\,I^A_{3,7}~,\nonumber\\
  I[1,3,4,6] =& I^C_{1,2,6,9}, \qquad
  I[2,3,5,6] = I^B_{1,2,3,8}, \qquad
  I[1,2,4,5] = I^D_{3,8,11,12} ~,
\end{align}
for the RR master integrals, and 
\begin{align}
    V[1,3] &= I^E_{4,5}~, &
    V[1,4] &= I^E_{5,7}~, &
    V[2,4] &= I^F_{4,12}~,\nonumber\\
    V[1,3,4] &= I^E_{2,5,7}~,&
    V[1,2,3,4,5] &= I^G_{1,2,8,9,10}~,&
    C[1,2,3,4] &= I^E_{2,4,5,7}~,
\end{align}
for the RV master integrals.

To compute the master integrals, we build a system of differential equations with respect to $z$ for the master integrals and bring it into canonical form~\cite{Henn:2013pwa} using \texttt{Fuchsia}~\cite{Gituliar:2017vzm}. We solve the system with iterated integrals, where we notice that the only letters appearing are the ones related to HPLs:
\begin{equation}
    \mathcal{A}_z =\left\{ \frac{1}{z},\frac{1}{1 - z},\frac{1}{1 + z}\right\}. 
\end{equation}
The solution contains yet undetermined boundary conditions for all master integrals. 

\subsubsection{Required boundary conditions}
\label{subsec:minimization}
A priori, one boundary condition per master integral must be determined from a calculation of the integral at some fixed value of $z$. However, for many of the master integrals, this boundary condition can be inferred from the self-consistency of the solution of the differential equation in the limit $z\to 1$. 

Expanding the differential equations around $z=1$, one finds that all RR master integrals must behave as 
\begin{equation}
    I_i^{RR} \sim (1-z)^{n_i - 2 \epsilon}, n_i\in \mathbb{Z}
    ~,
    \label{eq:scalRR}
\end{equation}
where $n_i$ is a constant that can be determined for each integral and depends on the soft scaling of its propagators. In the case of the RV master integrals the loop momentum is not constrained to the soft scaling. Therefore, we find that in the limit $z \to 1$ they can develop two possible branches
\begin{align}
    I_i^{RV} \sim I_i^{RV,(1)} (1-z)^{m_i - \epsilon} + I_i^{RV,(2)} (1-z)^{n_i - 2\epsilon} 
    ~.
    \label{eq:scalRV}
\end{align}

For the RR master integrals we can factor out the uniform scaling behaviour and impose the cancellation of all the terms proportional to
\begin{equation}
    (1-z)^{-i}, \log^i({1-z}) \quad i\geq 1
\end{equation}
in the limit $z\rightarrow 1$.
This yields a system of equations for the boundary terms at $z=1$, which leads to a substantial reduction of the unknown boundary terms. 
After solving this system, we find that the only RR integrals for which we need to compute the boundaries are
\begin{equation}
    I_0, I^A_{1,7}, I^B_{1,2,3,8},I^C_{1,2,6,9}.
    \label{eq:mitocompute}
\end{equation}
We note that only two of them are  top-sector integrals, from integral families B and C. 

For the RV masters we separate the discussion into the two branches discussed in Eq.~\eqref{eq:scalRV}. 
Since the determination of the leading behaviour is more involved in the RV case, we only
divide out the $\epsilon$-dependent scaling in the limit $z \to 1$ for the individual branches
and demand the cancellation of all logarithms in $1-z$. 
For the first branch, we find that we have to compute the following boundaries
\begin{align}
I^{E,(1)}_{4,5},I^{E,(1)}_{5,7}
~,
\end{align}
and for the second branch
\begin{align} 
I^{F,(2)}_{4,12},I^{G,(2)}_{1,2,8,9,10}
~.
\end{align}

We calculate the boundary conditions for (\ref{eq:mitocompute}) in the soft limit $z\rightarrow 1$, by parametrizing the integrals as follows
\begin{align}
    I_i^{RR} &\sim (1-z)^{n_i - 2 \epsilon} \sum_j c_j(\epsilon) (1-z)^j 
    ~, \\
    I_i^{RV,(1)} &\sim (1-z)^{m_i - \epsilon} \sum_j d_j(\epsilon) (1-z)^j 
    ~, \\
    I_i^{RV,(2)} &\sim (1-z)^{l_i - 2\epsilon} \sum_j e_j(\epsilon) (1-z)^j 
    ~.
\end{align}
The boundary conditions to the differential equations are fully contained in the leading coefficients $c_0(\epsilon)$, $d_0(\epsilon)$ or 
$e_0(\epsilon)$ for each integral. These will be computed below by employing the AAMFlow method.

\subsection{Boundary terms for RR master integrals}

\subsubsection{Set-up of the analytic AMFlow}
\begin{figure}
     \centering
     \begin{subfigure}[b]{0.3\textwidth}
         \centering
         \includegraphics[width=\textwidth]{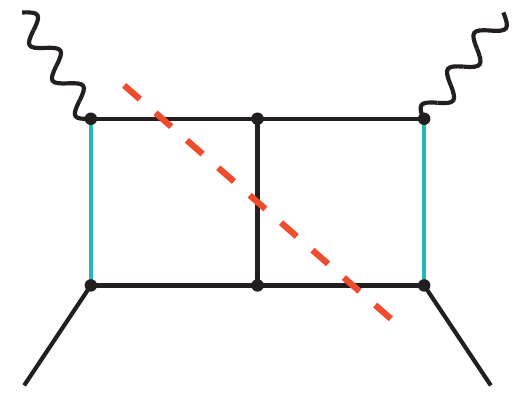}
         \caption{Corner integral of $\text{B}^{\text{aux}}$.}
         \label{fig:fam B aux}
     \end{subfigure}
     \hspace{0.1\textwidth}
     \begin{subfigure}[b]{0.3\textwidth}
         \centering
         \includegraphics[width=\textwidth]{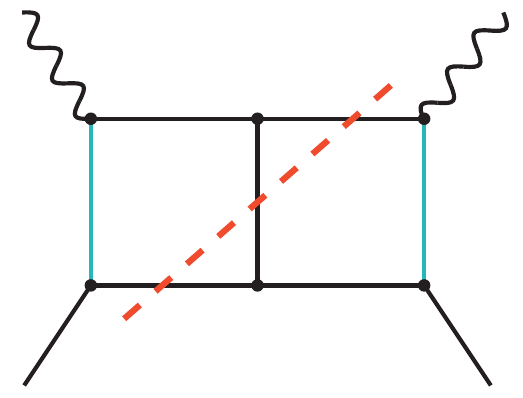}
         \caption{Corner integral of $\text{C}^{\text{aux}}$.}
         \label{fig:fam C}
     \end{subfigure}
        \caption{Corner integrals of the auxiliary families.}
        \label{fig:auxfams}
\end{figure}

Following the prescription presented in Sec.~\ref{sec:intro-amflow}, we create new auxiliary integral families for the original families B and C by adding an auxiliary mass to certain propagators. We define the following auxiliary families, illustrated by their corner integrals in Figure~\ref{fig:auxfams}:
\begin{align}
    \text{B}^{\text{aux}}&=\left\{
 {D_1^{\eta}}, {D_2^{\eta}}, D_3, \slashed{D}_6, D_8, \slashed{D}_9, \slashed{D}_{10} \right\} \,,\nonumber\\
    \text{C}^{\text{aux}}&=\left\{{D_1^{\eta}}, {D_2^{\eta}}, \slashed{D}_3, D_6, \slashed{D}_8, D_9, \slashed{D}_{10}\right\}\,.
\end{align}
where we indicate the propagators with an auxiliary mass as
\begin{equation}
    {D_i^{\eta}} \equiv D_i - \eta^2.
\end{equation} 

We separately reduce each family to master integrals using~\texttt{Kira}, without mapping the two via symmetry relations. We obtain $8$ master integrals for family $\text{B}^{\text{aux}}$ and $5$ for family $\text{C}^{\text{aux}}$. These families are the only ones we need to calculate for the boundaries, since $I^A_{1,7}$ is mapped via symmetry relations to $I^B_{1,2}$. For each auxiliary family, we build a system of differential equations with respect to the new variables $u=1/\eta^2$ and $y=1-z$, as these are more convenient for fixing the constants of integration in the large mass limit $\eta^2\to \infty$ ($u\rightarrow 0$). 

The physical value of the master integrals around $z=1$ is then obtained by solving the differential equations in $u$ in terms of iterated integrals, retaining only the leading term in small $y$, followed by taking the limit of vanishing auxiliary mass $\eta \to 0$ ($u\to \infty$). This implies performing a change of variables from $u$ back to $\eta^2$, which includes an analytic continuation of the iterated integrals. 

\subsubsection{Family B}
\label{subsec:famB}
We set up a differential equation with respect to $u=1/\eta^2$ for the integrals in family $\text{B}^{\text{aux}}$. 
The master integrals are:
\begin{align}
    &I_{0} = \int \frac{1}{\slashed{D}_6 \slashed{D}_9 \slashed{D}_{10}} \nonumber\\
    &B^{\text{aux}}_{1} = \int \frac{1}{{D_1^{\eta}} \slashed{D}_6 \slashed{D}_9 \slashed{D}_{10}} \nonumber\\
    &B^{\text{aux}}_{3/1} = \int \frac{D_3}{{D_1^{\eta}} \slashed{D}_6 \slashed{D}_9 \slashed{D}_{10}} \nonumber\\
    &B^{\text{aux}}_{1,2} = \int \frac{1}{{D_1^{\eta}} {D_2^{\eta}} \slashed{D}_6 \slashed{D}_9 \slashed{D}_{10}} \nonumber\\
    &B^{\text{aux}}_{3/1,2} = \int \frac{D_3}{{D_1^{\eta}} {D_2^{\eta}} \slashed{D}_6 \slashed{D}_9 \slashed{D}_{10}} \nonumber\\
    &B^{\text{aux}}_{3^2/1,2} = \int \frac{D_3^2}{{D_1^{\eta}} {D_2^{\eta}} \slashed{D}_6 \slashed{D}_9 \slashed{D}_{10}} \nonumber\\
    &B^{\text{aux}}_{2,3} = \int \frac{1}{{D_2^{\eta}} D_3 \slashed{D}_6 \slashed{D}_9 \slashed{D}_{10}} \nonumber\\
    &B^{\text{aux}}_{1,2,3,8} = \int \frac{1}{{D_1^{\eta}} {D_2^{\eta}} D_3 \slashed{D}_6 D_8 \slashed{D}_9 \slashed{D}_{10} }.
    \label{eq:misBaux}
\end{align}

Performing the same scaling analysis we did in Sect.~\ref{subsec:minimization}, we find the following behaviour in the limit $y \to 0$
\begin{align}
    I_0 &\rightarrow y^{1-2\epsilon} ,& B^{\text{aux}}_1 &\rightarrow y^{1-2\epsilon} ,& B^{\text{aux}}_{3/1} &\rightarrow y^{2-2\epsilon} ,& B^{\text{aux}}_{1,2} &\rightarrow y^{1-2\epsilon} 
    \nonumber \\
     B^{\text{aux}}_{3/1,2} &\rightarrow y^{2-2\epsilon} ,& B^{\text{aux}}_{3^2/1,2} &\rightarrow y^{3-2\epsilon} ,& B^{\text{aux}}_{2,3} &\rightarrow  y^{-2\epsilon} ,& B^{\text{aux}}_{1,2,3,8} &\rightarrow y^{-1-2\epsilon}
     ~.
\end{align}
We need the leading term of $\mathcal{O}(y^{-1-2\epsilon})$ of the top sector to calculate the boundary condition for the physical integral $I^B_{1,2,3,8}$. Therefore we only need to solve the differential equation associated to the top sector up to this order in $y$. It turns out that at this order, the inhomogeneous term of the differential equation for $B^{\text{aux}}_{1,2,3,8}$ receives contributions only from the integrals $B_0$, $B_1^{\text{aux}}$, $B^{\text{aux}}_{1,2}$ and $B^{\text{aux}}_{2,3}$. The differential equations for these integrals do however involve all other masters of the $B^{\text{aux}}$ topology. We give details on the calculation in the following.

The integrals $B_1^{\text{aux}}$ and $B_{3/1}^{\text{aux}}$ form a coupled system of differential equations:
\begin{align}
\partial_u B^{\text{aux}}_1&= \frac{\left(2 \epsilon u^2-4 \epsilon u y+5 \epsilon u-3 \epsilon y+3 \epsilon-u^2+2 u y-2 u+y-1\right)}{u (u+1) (u-y+1)}\, B^{\text{aux}}_1  \nonumber\\
&-\frac{(3 \epsilon-2) (y-1) }{u-y+1}\, I_0 +\frac{(3 \epsilon-2) u (y-1) }{(u+1) (u-y+1)}\, B^{\text{aux}}_{3/1},\nonumber\\
\partial_u B^{\text{aux}}_{3/1}&= \frac{(\epsilon-1) y}{u (u+1) (y-1)}\, B^{\text{aux}}_1 +\frac{(3 \epsilon-2) }{u (u+1)}\, B^{\text{aux}}_{3/1}.
\end{align}
We decouple it into a second-order differential equation for $B_1^{\text{aux}}$,
\begin{align}
    \partial^2_u B^{\text{aux}}_1&=\left(-\frac{2 \epsilon}{u-y+1}-\frac{2 \epsilon}{u+1}+\frac{6 \epsilon-2}{u}\right)\,\partial_u B^{\text{aux}}_1 \nonumber\\
    &+\frac{\epsilon (3 \epsilon-1) ((u+3) y-2 u-3)}{u^2 (u+1) (u-y+1)}\,B^{\text{aux}}_1 \nonumber\\
    &+\frac{(9 (\epsilon-1) \epsilon+2) (y-1)}{u (u+1) (u-y+1)}\, I_0 ,
\end{align}
and an algebraic relation for $B^{\text{aux}}_{3/1}$.

We solve the second order differential equation by inserting a symbolic ansatz:
\begin{equation}
    B_1^{\text{aux}} \rightarrow \sum_{i=1}^2\sum_{j=-4}^6 c_{i j}(u) y^i \epsilon^j,
\end{equation}
and solving for the $c_{ij}(u)$, up to integration constants. The large-mass limit of $B_1^{\text{aux}}$ is given by
\begin{equation}
   \lim_{u\to 0} \, \frac{1}{u} B_1^{\text{aux}} = - I_0,
\end{equation}
and allows us to fix all the remaining constants of integration. 

The master integrals $B^{\text{aux}}_{1,2}, B^{\text{aux}}_{3/1,2}, B^{\text{aux}}_{3^2/1,2}$ fulfil a coupled $ 3\cross 3$ system of inhomogeous differential equations. We decouple this system to get a third order differential equation for $B^{\text{aux}}_{1,2}$ using \texttt{OreSys}~\cite{oresys} (which is based on \texttt{Sigma}~\cite{sigmaI,sigmaII}). We solve this differential equation using~\texttt{HarmonicSums}~\cite{Vermaseren:1998uu,Blumlein:1998if,Remiddi:1999ew,Blumlein:2003gb,Blumlein:2009ta,Ablinger:2009ovq,Ablinger:2011te,Ablinger:2012ufz,Ablinger:2013eba,Ablinger:2013cf,Ablinger:2014rba,Ablinger:2014bra,Ablinger:2015gdg,Ablinger:2018cja}
and fix its constants of integration with the large-mass expansion of $B^{\text{aux}}_{1,2}$
\begin{equation}
    \lim_{u\to 0} \, \frac{1}{u^2} B_{1,2}^{\text{aux}} =  I_0,
\end{equation}

The remaining integral 
$B^{\text{aux}}_{2,3}$ fulfils a simple first-order 
inhomogeneous differential equation:
\begin{align}
    \partial_u B^{\text{aux}}_{2,3}&=\frac{(3 \epsilon-2) (2 \epsilon-1) (y-1)^2 }{\epsilon y (u-y+1)}I_0-\frac{(2 \epsilon-1)^2 (y-1) (u-2 y+1) }{\epsilon (u+1) y (u-y+1)}B^{\text{aux}}_1\nonumber\\
    &-\frac{(3 \epsilon-2) (2 \epsilon-1) u (y-1)^2 }{\epsilon (u+1) y (u-y+1)}B^{\text{aux}}_{3/1}+\frac{2 \epsilon}{u (u+1)} B^{\text{aux}}_{2,3}.
\end{align}
We insert a symbolic ansatz for $B^{\text{aux}}_{2,3}$ that respects its scaling in $y$ ($B^{\text{aux}}_{2,3} \sim y^{0-2\epsilon}$):
\begin{equation}
    B^{\text{aux}}_{2,3} \rightarrow y^{-2\epsilon} \sum_{i=-4}^5 \epsilon^i c_i(u)
\end{equation}
and we solve the differential equations for the $c_i(u)$ order by order in $\epsilon$.
The remaining constants of integration 
are fixed from the large-mass behaviour
\begin{equation}
    \lim_{u \to 0}\, \frac{1}{u} B^{\text{aux}}_{2,3} =  B^{\text{aux}}_{3} = u \frac{(3 \epsilon-2) (y-1)}{\epsilon y}\, I_0.
\end{equation}

We can now consider the differential equation for the top sector. From the scaling in $y$ of the physical top sector, we know that it is proportional to $y^{-1-2\epsilon}$ and therefore we require only that order in the expansion of its differential equation in order to determine the leading term of its $y$ expansion. Taking into account the $y$-scaling of all subsector master integrals we then expand the top-sector differential equation for $y\rightarrow 0$ and keeping only the term of $O(y^{-1-2\epsilon})$:
\begin{align}
    \partial_u B^{\text{aux}}_{1,2,3,8}&= \frac{2 (2 \epsilon-1) (3 \epsilon-2) u (2 \epsilon u-3 \epsilon+1)}{\epsilon^2 (u+1) (u+2) y^2}\,I_0 \nonumber\\
    &+\frac{4 (2 \epsilon-1)^2 u^2}{\epsilon (u+1) (u+2) y^2}\, B^{\text{aux}}_{1}+\frac{4 (2 \epsilon-1) u }{(u+1) (u+2) y^2}\, B^{\text{aux}}_{1,2}\nonumber\\ 
    &+\frac{4 \epsilon u }{(u+1) (u+2) y}\,B^{\text{aux}}_{2,3}+\frac{2 (2 \epsilon+1)}{u (u+2) } \,B^{\text{aux}}_{1,2,3,8} + \mathcal{O}(y^0).
    \label{eq:DEtopsecBaux}
\end{align}
Then, we insert a symbolic $\epsilon$-expansion for each integral and build a differential equation system for each $\epsilon$-order into which the solutions for all the subsectors are inserted. The resulting differential equations are solved using \texttt{HarmonicSums}. The large-mass expansion of the top sector is given by 
\begin{equation}
   \lim_{u\rightarrow 0} \, \frac{1}{u^2} B^{\text{aux}}_{1,2,3,8} = B^{\text{aux}}_{3,8},
\end{equation}
which is a subsector integral not reducible to the phase-space. However, for the purpose of fixing the boundary values it is sufficient to impose that the solution scales as $u^2$.

After fixing the constants of integration we find
\begin{align}
    B^{\text{aux}}_{1,2,3,8} &= 
    S_\Gamma \pi \,
    y^{-1+2\epsilon}
    \frac{2 u}{2+u} 
    \biggl\{ 
    \frac{1}{\epsilon^2}
    \biggl(
        2 H_{-1}(u)
    \biggr)
    + \frac{1}{\epsilon} 
    \biggl( 
          4  H_{0,-1}(u)
        + 4 H_{-1,-1}(u)
        - 4 H_{-2,-1}(u)
    \biggr)
    \nonumber \\ &
        +  8 H_{0,0,-1}(u)
        + 8 H_{0,-1,-1}(u)
        - 8 H_{0,-2,-1}(u)
        + 16 H_{-1,0,-1}(u)
        \nonumber \\ &
        - 16 H_{-1,-1,-1}(u)
        + 8 H_{-1,-2,-1}(u)
        - 8 H_{-2,0,-1}(u)
        - 8 H_{-2,-1,-1}(u)
        \nonumber \\ &
        + 8 H_{-2,-2,-1}(u)
        - 10 \zeta_2 H_{-1}(u) 
    + \mathcal{O}(y^1,\epsilon) 
    \biggr\}
    ~.
\end{align}
The result has been obtained up to and including terms of $\mathcal{O}\left(\epsilon^5\right)$. The result is given by iterated integrals over the alphabet 
\begin{align} 
    \mathcal{A}_u &= \biggl\{ w_0=\frac{1}{x} , w_{-1}=\frac{1}{1+x}, w_{-2}=\frac{1}{2+x} \biggr\} 
    ~.
\end{align}
In the final step, the analytic continuation to $\eta = 1/u \to 0$ has to be performed. The analytic continuation and subsequent expansion in $\eta$ can be performed with routines implemented in \texttt{HarmonicSums}, e.g.\ we find 
\begin{align}
    H_{-2,0,-1}(u) &= 
     -H_{-\frac{1}{2},0,-1}(\eta^2)
     -H_{0,0,0}(\eta^2)
     +H_{-\frac{1}{2}}(\eta^2) \zeta_2
     -H_0(\eta^2) \zeta_2
     +H_{-\frac{1}{2},0,0}(\eta^2)
     \nonumber \\ &
     +H_{0,0,-1}(\eta^2)
     -\frac{3}{4} \zeta_3
     - \zeta_2 H_{-\frac{1}{2}}(1)
    +H_{-\frac{1}{2},0,-1}(1)
    -H_{-\frac{1}{2},0,0}(1)
    \nonumber \\ &
    +H_{-2,0,-1}(1)
    \\
    &= 
    -H_{0,0,0}(\eta^2)
     -H_0(\eta^2) \zeta_2
     -\frac{3}{4} \zeta_3
     - \zeta_2 H_{-\frac{1}{2}}(1)
    +H_{-\frac{1}{2},0,-1}(1)
    \nonumber \\ &
    -H_{-\frac{1}{2},0,0}(1)
    +H_{-2,0,-1}(1)
    +\mathcal{O}\left(\eta^2\right).
\end{align}
Due to the analytic continuation, we are left with constants given by iterated integrals over the alphabet 
\begin{align}
    \mathcal{A}_{\eta} &= \biggl\{ w_0=\frac{1}{x} , w_{-1}=\frac{1}{1+x}, w_{-2}=\frac{1}{2+x} , w_{-1/2}=\frac{1}{1/2+x} \biggr\}
    \label{eq:alphabeteta}
\end{align}
evaluated at argument 1.

As long as only the first two letters of the alphabet in Eq.~\eqref{eq:alphabeteta} contribute, the constants are well known and reduction tables up to high weight have been computed~\cite{Remiddi:1999ew,Blumlein:2009cf}.
Once the other letters of the alphabet contribute, we do not have such a complete reduction of constants. A practical way to deal with the new kind of constants appearing is to evaluate them with high precision utilizing for example {\tt ginac}~\cite{Vollinga:2004sn} and reconstruct the analytic structure using the {\tt PSLQ} algorithm~\cite{pslq}. This procedure will succeed since we know that for the physical branch in the limit $z \to 1$ only multiple $\zeta$ values contribute.

Alternatively, the constants can also be reduced analytically by replacing the integer $2$ in the alphabet \eqref{eq:alphabeteta} by a parameter $t$ and performing a subsequent fibration to HPLs of argument $t$. The resulting HPLs at argument $t=2$ are then evaluated in terms of known transcendental constants.  

We find for the leading expansion term in the limit $\eta \to 0$
\begin{align*}
    B^{\text{aux}}_{1,2,3,8}
    &= 
    S_\Gamma \pi \,
    y^{-1+2\epsilon} 
    \biggl\{ 
        \frac{-4 H_0(\eta^2)}{\epsilon^2}
        +\frac{1}{\epsilon} \biggl( 4 H_0^2(\eta^2) + 2\pi^2 \biggr)
        - \frac{8}{3} H_0^3(\eta^2) - 2\frac{\pi^2}{3} H_0(\eta^2)  
        \nonumber \\ &
        + 16 \zeta_3
        + \mathcal{O}(\epsilon)
    \biggr\}
    +\mathcal{O}\left(y^0\right)
    ~,
\end{align*}
where we again only give explicit results up to $\mathcal{O}(\epsilon^0)$.

We extract the hard region according to Eq.~\eqref{eq:fix}. As a first system of equations we find
(where we suppress the global factor of $S_\Gamma \pi$): 
\begin{align}
\begin{cases}
    -d_{ij,1}^{-3}-2 d_{ij,2}^{-3} &= -4~,\\
    d_{ij,1}^{-3} / 2+2 d_{ij,2}^{-3} &= \hphantom{-}4~,\\
    d_{ij,0}^{-3} + d_{ij,1}^{-3} + d_{ij,2}^{-3} &= \hphantom{-}0 ~,
\end{cases}
\end{align}
with the solution 
\begin{align*} 
    d_{ij,2}^{-3} &= 0 ~, 
    &
    d_{ij,1}^{-3} &= 2 ~,
    &
    d_{ij,0}^{-3} &= -2.
\end{align*}
It turns out that the branch with $n=2$ does not contribute in general for this integral. Proceeding like this also for the higher orders in $\epsilon$, we finally obtain the physical boundary condition 
\begin{align}
    I^B_{1,2,3,8}\eval_{y\rightarrow 0} &= 
    S_\Gamma \pi \,
    y^{-1+2\epsilon}
    \biggl\{ 
    - \frac{2}{\epsilon^3}
    + \frac{5 \pi^2}{3 \epsilon}
    + \frac{76 \zeta_3}{3}
    + \frac{7 \pi^4}{36} \epsilon 
    + \biggl( \frac{1124\zeta_5}{5} -\frac{148 \pi^2 \zeta_3}{9} \biggr) \epsilon^2
    \nonumber \\ &
    + \biggl( \frac{155 \pi^6}{504} - \frac{382 \zeta_3^2}{9} \biggr) \epsilon^3
    + \mathcal{O}(\epsilon^4)
    \biggr\}
    + \mathcal{O}(y^0) 
    ~.
\end{align}
We notice that the boundary is of uniform transcendental weight.

\subsubsection{Family C}
For the auxiliary family C we find the following master integrals:
\begin{align}
    &I_{0} = \int \frac{1}{\slashed{D}_3 \slashed{D}_8 \slashed{D}_{10}} \nonumber\\
    &C^{\text{aux}}_{1} = \int \frac{1}{{D_1^{\eta}} \slashed{D}_3 \slashed{D}_8 \slashed{D}_{10}} \nonumber\\
    &C^{\text{aux}}_{1,2} = \int \frac{1}{{D_1^{\eta}} {D_2^{\eta}} \slashed{D}_3 \slashed{D}_8 \slashed{D}_{10}} \nonumber\\
    &C^{\text{aux}}_{2,6} = \int \frac{1}{{D_2^{\eta}} \slashed{D}_3 D_6 \slashed{D}_8 \slashed{D}_{10}} \nonumber\\
    &C^{\text{aux}}_{1,2,6,9} = \int \frac{1}{{D_1^{\eta}} {D_2^{\eta}} \slashed{D}_3 D_6 \slashed{D}_8 D_9 \slashed{D}_{10}}.
    \label{eq:famCMI}
\end{align}

Performing the same scaling analysis we did in Sect.~\ref{subsec:minimization}, we find the following behaviour in the limit $y \to 0$ 
\begin{align}
    I_0 &\rightarrow y^{1-2\epsilon} ,& 
    C^{\text{aux}}_1 &\rightarrow y^{1-2\epsilon} ,& 
    C^{\text{aux}}_{1,2} &\rightarrow y^{1-2\epsilon} ,\nonumber\\
    C^{\text{aux}}_{2,6} &\rightarrow y^{-2\epsilon},
     &C^{\text{aux}}_{1,2,6,9} &\rightarrow y^{-1-2\epsilon} 
     ~.
\label{eq:Cauxyscaling}
\end{align}

The differential equations in auxiliary family C are not coupled and we can simply 
solve them sequentially in a bottom-up approach. Starting from $C^{\text{aux}}_1$, for each subsector integral we solve the differential equation with respect to $u$ and fix the constants of integration in the large mass limit. We notice that there is one subsector integral, $C^{\text{aux}}_{2,6}$, for which the large mass limit does not correspond to the phase space. Its constants of integration are however easily fixed by imposing the correct scaling in $u$ of the solution.

Having the solution for all the subsectors, we can turn to the differential equation for the top sector $C^{\text{aux}}_{1,2,6,9}$, which is the one we actually need for the physical 
boundary conditions.
Its differential equation reads
\begin{align}
   \partial_u C^{\text{aux}}_{1,2,6,9}&=
   -\frac{2 (2 \epsilon-1) (3 \epsilon-2) u (y-1)^4 (5 \epsilon u-3 \epsilon y+3 \epsilon-u+y-1)}{\epsilon^2 y^2 (u-2 y+2) (u-y+1)^2}\, I_0 \nonumber\\
   &-\frac{4 (1-2 \epsilon)^2 u^2 (y-1)^3 }{\epsilon y^2 (u-2 y+2) (u-y+1)^2}\,C^{\text{aux}}_{1} \nonumber\\
   &-\frac{4 (1-2 \epsilon) u (y-1)^2 }{y^2 (u-2 y+2) (u-y+1)}\,C^{\text{aux}}_{1,2} \nonumber\\
   &-\frac{4 \epsilon u (y-1)^2 }{y (u-2 y+2) (u-y+1)}\, C^{\text{aux}}_{2,6} +\frac{2 (-2 \epsilon-1) (y-1) }{u (u-2 y+2)}\, C^{\text{aux}}_{1,2,6,9}.
\end{align}
To have a smooth workflow, we first substitute a symbolic ansatz for each master integral (except for $I_0$), where the leading order in  $y$ is determined in eq.~\eqref{eq:Cauxyscaling},
and expand the obtained system up to the $y$-order needed to fix the leading order of the top sector, which is 
of $\mathcal{O}(y^{-1-2\epsilon})$. We then insert the actual expression of the coefficients. The differential equation is solved using \texttt{HarmonicSums} and the algorithm presented in~\cite{Ablinger:2018zwz}. The large-mass expansion for $C^{\text{aux}}_{1,2,6,9}$ reads
\begin{equation}
    \lim_{u\to 0}\, \frac{1}{u^2} C^{\text{aux}}_{1,2,6,9} = C^{\text{aux}}_{6,9} =
    \frac{\left(9 \epsilon^2-9 \epsilon+2\right) (y-1)^2}{\epsilon^2 y^2}\, I_0.
\end{equation}

Also in this case  all  constants of integration could be fixed by requiring that the terms of order $u^i, i\leq 1$ vanish, without explicitly calculating $C^{\text{aux}}_{6,9}$. 
Following the same steps presented in Sect.~\ref{subsec:famB}, we analytically continue our solution to the variable $\eta^2$ and 
take the limit $\eta^2 \to 0$. 
Extracting the hard-region of this limit allows us to calculate the first coefficient of the $y$ expansion of $I^C_{1,2,6,9}$: 
\begin{align} 
    I^C_{1,2,6,9}\eval_{y \to 0}
    &= 
    S_\Gamma \pi \,
    y^{-1+2\epsilon} 
    \biggl\{ 
    -\frac{10}{\epsilon^3}
    +\frac{25 \pi ^2}{3 \epsilon}
    -\frac{55 \pi ^4 \epsilon}{36}
    +\frac{296 \zeta_3}{3} 
    +\epsilon^2 \left(256 \zeta_5 -\frac{740 \pi ^2 \zeta_3}{9}\right)\nonumber\\
    &+\epsilon^3 
    \left(
        -\frac{4640 \zeta_3^2}{9}
        -\frac{215 \pi ^6}{504}\right)
        + \mathcal{O}(\epsilon^4)
    \biggr\}
    +\mathcal{O}\left(y^0\right),
\end{align}
thereby providing the boundary condition in the limit $z \to 1$ for $I^C_{1,2,6,9}$.

\subsection{Boundary terms for RV master integrals}
The procedure outlined above for the boundary conditions to the RR master integrals can also be applied in the case of the RV master integrals. However, since masses are now also added to loop-momentum propagators, the asymptotic expansion for $\eta \to \infty$ gets more involved.

We illustrate this here on the example of fixing the boundary conditions for the top-sector RV master integral $I^{G,(2)}_{1,2,8,9,10}$. It is noted that the other RV master integrals $I^{E}_{4,5}$, $I^{E}_{5,7}$ and $I^{F}_{4,12}$ all correspond to one-loop bubble graphs, whose boundary conditions are obtained in a straightforward manner by direct integration. This example allows us to document the more complicated structures arising when applying our workflow to loop integrals.

\subsubsection{Set-up of the analytic AMFlow}
\begin{figure}
         \centering
         \includegraphics[width=0.3\textwidth]{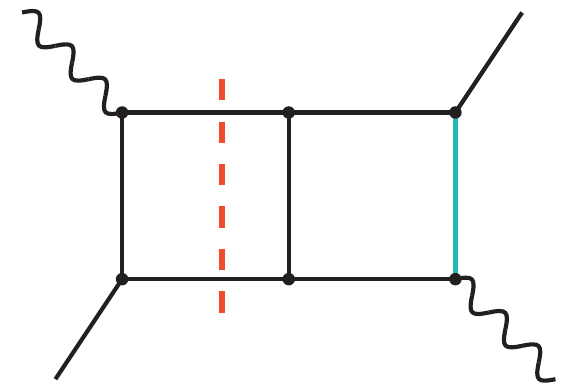}
         \caption{Corner integral of $\text{G}^{\text{aux}}$.}
         \label{fig:fam G aux}
\end{figure}

We add the auxiliary mass only to propagator $D_2$, as depicted in Fig.~\ref{fig:fam G aux}. This minimal choice allows to limit the growth in the number of master integrals and makes the $y$ and $\eta^2$ expansion independent. In the following, we indicate this auxiliary family as $G^{\text{aux}}$:
\begin{align}
    \text{G}^{\text{aux}}&=\left\{D_1, {D_2^{\eta}}, \slashed{D}_3,\slashed{D}_6, D_8, D_9,D_{10}\right\}\,.
\end{align}
We find nine master integrals:
\begin{align}
    &G^{\text{aux}}_2 = \int \frac{1}{{D_2^{\eta}} \slashed{D}_3 \slashed{D}_6} \nonumber\\
    &G^{\text{aux}}_{2,9} = \int \frac{1}{D_2^{\eta} \slashed{D}_3 \slashed{D}_6 D_9}\nonumber\\
    &G^{\text{aux}}_{8,9} = \int \frac{1}{\slashed{D}_3 \slashed{D}_6 D_8 D_9} \nonumber\\
    &G^{\text{aux}}_{2,10} = \int \frac{1}{D_2^{\eta} \slashed{D}_3 \slashed{D}_6 D_{10}} \nonumber\\
    &G^{\text{aux}}_{9/2,10} = \int \frac{D_9}{D_2^{\eta} \slashed{D}_3 \slashed{D}_6 D_{10}} \nonumber\\
    &G^{\text{aux}}_{2,8,9} = \int \frac{1}{D_2^{\eta} \slashed{D}_3 \slashed{D}_6 D_{8} D_9}  \nonumber\\
    &G^{\text{aux}}_{2,8,10} = \int \frac{1}{D_2^{\eta} \slashed{D}_3 \slashed{D}_6 D_{8} D_{10}}  \nonumber\\
    &G^{\text{aux}}_{2,9,10} = \int \frac{1}{D_2^{\eta} \slashed{D}_3 \slashed{D}_6 D_{9} D_{10}}  \nonumber\\
    &G^{\text{aux}}_{1,2,8,9,10} = \int \frac{1}{D_1 D_2^{\eta} \slashed{D}_3 \slashed{D}_6 D_8 D_{9} D_{10}}.
    \label{eq:misGaux}
\end{align}
We know that these integrals contain two branches in the variable $y$, given by the different scalings of the loop momentum $k$. We therefore can write each master integral as the sum of two contributions: 
\begin{equation}
    G^{\text{aux}}_i = y^{-\epsilon} G^{\text{aux},(1)}_{i}\left(y,\epsilon,\eta^2\right) + y^{-2\epsilon} G^{\text{aux},(2)}_{i}\left(y,\epsilon,\eta^2\right).
    \label{eq:RV-y-exp}
\end{equation}
We derive the differential equation for the master integrals with respect to $u=1/\eta^2$
\begin{equation}
    \partial_u G^{\text{aux}}_i = \mathbf{M} \, G^{\text{aux}}_i.
\end{equation}
By inserting the ansatz~\eqref{eq:RV-y-exp}, we split the system into two independent systems, one for each branch
in the asymptotic expansion:
\begin{equation}
    \begin{cases}
         \partial_u  G^{\text{aux},(1)}_{i} &= \mathbf{M}  \, G^{\text{aux},(1)}_{i} ~,\\
          \partial_u  G^{\text{aux},(2)}_{i} &= \mathbf{M} \, G^{\text{aux},(2)}_{i} ~,
    \end{cases}
\end{equation}
We can now exploit the independence of the $y$ expansion with respect to $\eta^2$ to simplify the differential equation for each branch. In fact, due to this independence, also the large mass limit ($\eta^2 \rightarrow \infty$) of the master integrals presents an analogous structure of the form~\eqref{eq:RV-y-exp} and has the advantage to be more easily calculable. In particular, if one of the two terms $G^{(1)}_i, G^{(2)}_i$ is equal to zero, then it is also zero for arbitrary values of $\eta^2$, by the independence of the two expansions. This allows to nullify specific entries in $\mathbf{M}$ for each branch of $y$. As a result, our system now reads
\begin{equation}
    \begin{cases}
         \partial_u  G^{\text{aux},(1)}_{i} &= \mathbf{M_1}  \, G^{\text{aux},(1)}_{i} ~,\\
          \partial_u  G^{\text{aux},(2)}_{i} &= \mathbf{M_2} \, G^{\text{aux},(2)}_{i} ~,
    \end{cases}
\end{equation}
with $\mathbf{M_1}$ and $\mathbf{M_2}$ being 9$\times$9 matrices of coefficients. It is observed that $\mathbf{M_2}$ has non-vanishing entries only for $G^{\text{aux},(2)}_{8,9}$, $G^{\text{aux},(2)}_{2,8,9}$ and $G^{\text{aux},(2)}_{1,2,8,9,10}$. Consequently, the differential equation for the $y^{-2\epsilon}$ branch, the only one needed for the top-sector boundary, dramatically simplifies. 

The subsector integral $G^{\text{aux}}_{8,9}$ is independent of the auxiliary mass. In particular, it consists of the integration over the $2\to2$ particle phase space of a massless bubble. We can easily calculate it in closed form: 
\begin{align}
   G^{\text{aux}}_{8,9}= 2\pi i \, S_\Gamma \,e^{2 \gamma  \epsilon} y^{-2 \epsilon}  (1-y)^{\epsilon}  \cos( \pi \epsilon) \,\frac{ \Gamma (1-\epsilon)^3 \Gamma (\epsilon)}{\Gamma (2-2 \epsilon)^2}.
   \label{eq:gaux89}
\end{align}
We notice that the only $y$ branch contributing is $y^{-2 \epsilon}$, so~$G^{\text{aux}}_{8,9} =~G^{\text{aux},(2)}_{8,9}$. The subsector integral $G^{\text{aux},(2)}_{2,8,9}$, satisfies the following differential equation 
\begin{align}
    \partial_u G^{\text{aux},(2)}_{2,8,9} = \frac{\epsilon (u-2 y+2) }{u (u-y+1)}\,G^{\text{aux},(2)}_{2,8,9} -\frac{(2 \epsilon-1) (y-1) }{u-y+1}\,G^{\text{aux}}_{8,9},
\end{align}
that can be easily solved by inserting $G^{\text{aux}}_{8,9}$~\eqref{eq:gaux89}. The integration constants are fixed in the limit of large auxiliary mass. 

As a general feature, when applying our workflow to one-loop integrals, we can distinguish two regions, where the loop momentum $k$ scales either as large $(L)$ as the auxiliary mass, $k \sim \mathcal{O}(\eta)$, or small $(S)$ with respect to it, $k \sim \mathcal{O}(1)$. The sum of the two regions gives the full large-mass expansion: 
\begin{align}
    \lim_{u\to 0}G^{\text{aux}} = \lim_{u\to 0, k \sim \text{small}}G^{\text{aux}}+ \lim_{u\to 0, k \sim \text{large}} G^{\text{aux}} = G^{\text{aux},(L)} + G^{\text{aux},(S)}.
\end{align}

In this particular case, since we are interested in the $y^{-2 \epsilon}$ branch of the large-mass expansion, it is sufficient to examine only the small $(S)$ region. The reason is the following: the integration over the $2\to 2$ particle phase space produces a factor $y^{-\epsilon}$. In the large momentum region, all integrals over the $2\to 2$ particle phase space contain massive tadpoles, which do not present any $y$ dependence, and therefore can not give any extra factor of $y^{-\epsilon}$. For those integrals where it is non-vanishing, the large-mass expansion for the $y^{-2\epsilon}$ branch must therefore arise from the small $(S)$ region.

We find the following $u\to 0$ behaviour for $G^{\text{aux},(2)}_{2,8,9}$: 
\begin{align}
    G^{\text{aux},(S),(2)}_{2,8,9} &= -u\,G^{\text{aux},(S)}_{8,9}.
\end{align}
We fix the constants of integration as explained in Sect.~\ref{sec:intro-amflow}. The differential equation for the $y^{-2\epsilon}$ branch of $G^{\text{aux}}_{1,2,8,9,10}$ is
\begin{align}
    \partial_u G^{\text{aux},(2)}_{1,2,8,9,10} &= -\frac{2 (2 \epsilon -1)^2 (y-1)^3 }{\epsilon  y (u-y+1)}\,G^{\text{aux}}_{8,9} +\frac{2 (2 \epsilon -1) (y-1)^2 }{y (u-y+1)}\, G^{\text{aux},(2)}_{2,8,9}\nonumber\\
    &-\frac{(2 \epsilon +1) (y-1) }{u (u-y+1)}\, G^{\text{aux},(2)}_{1,2,8,9,10}
\end{align}
and can be solved after inserting in the analytic result for $G^{\text{aux}}_{8,9}$ and the solution for $G^{\text{aux}}_{2,8,9}$. 
The large-mass small-momentum expansion is given by 
     \begin{align}
    G^{\text{aux},(S),(2)}_{1,2,8,9,10} = -u\,G^{\text{aux},(S),(2)}_{1,8,9,10}= 
     -2\pi i \, S_\Gamma \, u \, y^{-1-2 \epsilon} \, e^{2 \gamma  \epsilon}\, \cos(\pi \epsilon)  \,\frac{  \Gamma (-\epsilon)^3 \Gamma (\epsilon+1)}{ \Gamma (1-2 \epsilon)^2}~,
\end{align}
which allows to fix the constants of integration as outlined in Sect.~\ref{sec:intro-amflow}. The leading term in the $y\to 0$ expansion of $I^{G,(2)}_{1,2,8,9,10}$ is then obtained as: 
\begin{align}
    I^{G,(2)}_{1,2,8,9,10}\eval_{y\to 0} &=
    2\pi i \, S_\Gamma \,
    y^{-1-2\epsilon} \biggl\{  
    -\frac{3}{\epsilon^3}
    +\frac{17 \pi ^2}{6 \epsilon}
    +24 \zeta_3
    -\frac{59 \pi ^4 \epsilon}{120} \nonumber\\
    &+\epsilon^2 \left(\frac{556 \zeta_5}{5}-\frac{188 \pi ^2 \zeta_3}{9}\right)
    +\epsilon^3 \left(-\frac{208 \zeta_3^2}{3}-\frac{463 \pi ^6}{45360}\right) \nonumber\\
    &+\epsilon^4 \left(\frac{193 \pi ^4 \zeta_3}{45}-\frac{474 \pi ^2 \zeta_5}{5}+\frac{3331 \zeta_7}{7}\right)
    +\mathcal{O}(\epsilon^5)\biggr\} +\mathcal{O}(y^0).
\end{align}
This integral provides the sought-for boundary condition to the differential equation of $I^{G,(2)}_{1,2,8,9,10}$.

\section{Conclusions}
\label{sec:conc}
The auxiliary mass flow method~\cite{Liu:2017jxz} has proven to be an immensely useful tool to determine loop and phase space master integrals for specific values of masses and kinematical parameters. It relies on differential equations in an auxiliary mass parameter, which are solved through high-precision series expansions, yielding its answers in purely numerical form that can in principle be obtained to a high number of desired digits. The method is often used in combination with the differential equation method~\cite{Kotikov:1990kg,Remiddi:1997ny,Gehrmann:1999as,Henn:2013pwa}, where it serves to provide the boundary conditions that are required to obtain the particular solutions for master integrals from the generic solutions of the differential equations.

In this work, we took up the idea of auxiliary mass flow to devise an analytic procedure to compute master integrals at specific kinematical points. The resulting Analytic Auxiliary Mass Flow (AAMFlow) method enables us to retain full control on the asymptotic structure of the solution in the physical limit, where the method is used to determine boundary conditions to differential equations. This feature allows in particular to compute boundary conditions at exceptional kinematical points, where master integrals often assume a particularly simple structure in terms of transcendental constants. 

To illustrate the new method, we computed phase space master integrals relevant to the NNLO corrections to deeply inelastic coefficient functions and to initial-final antenna functions to transcendental weight 6. Our results agree with previous derivations up to weight 4, and the novel terms will become relevant to extend the antenna subtraction method to N3LO. 
The analytic results are available in computer readable form as ancillary file to this publication.

By providing analytic expressions and allowing to address exceptional kinematical points, the AAMFlow method could in particular become relevant for few-scale loop or phase space master integrals, whose differential equations are amenable to the derivation of an analytical generic solution. By also providing the boundary terms in analytical form, the method will thus enable closed-form expressions for master integrals and amplitudes, which are ideally suited for mathematical investigations of their properties.

\section*{Acknowledgements}
We would like to thank Tiziano Peraro for useful discussions and feedback on the draft.
This work has received funding from the Swiss National Science Foundation (SNF) under contract 200020-204200 and from the European Research Council (ERC) under the European Union's Horizon 2020 research and innovation programme grant agreement 101019620 (ERC Advanced Grant TOPUP).
All diagrams have been made using~\texttt{JaxoDraw}~\cite{Binosi:2003yf}.

\appendix

\bibliographystyle{JHEP}
\bibliography{biblio.bib}

\end{document}